\documentclass[reprint,prl,aps,superscriptaddress,raggedbottom]{revtex4-2}
\usepackage{times}
\usepackage{graphicx}
\usepackage{amsmath,amssymb}
\usepackage[hidelinks,hyperfootnotes=false]{hyperref}
\newcommand\dif{\mathop{}\!\mathrm{d}}

\makeatletter
\def\maketitle{
\@author@finish
\title@column\titleblock@produce
\suppressfloats[t]}
\makeatother

\makeatletter
\let\frontmatter@footnote@produce\frontmatter@footnote@produce@footnote
\makeatother

\newcommand{\figureasset}[2]{%
  \IfFileExists{#1}{%
    \includegraphics[width=\linewidth,height=0.68\textheight,keepaspectratio]{#1}%
  }{%
    \fbox{\parbox[c][0.16\textheight][c]{0.94\linewidth}{\centering
      \textbf{#2}\\[4pt]\texttt{#1}}}%
  }%
}

\begin{document}

\title{Tunable first-order transitions and magnetic hysteresis in Bernal pentalayer graphene}

\newcommand{\BPhysics}{Department of Physics and Astronomy, Rice University, Houston, TX 77005, USA}

\newcommand{\BSCI}{Smalley-Curl Institute, Rice University, Houston, TX 77005, USA}

\newcommand{\BNIMSMaterials}{Research Center for Electronic and Optical Materials, National Institute for Material Science, 1-1 Namiki, Tsukuba 305-0044, Japan}

\newcommand{\BNIMSNano}{Research Center for Materials Nanoarchitectonics, National Institute for Material Science, 1-1 Namiki, Tsukuba 305-0044, Japan}

\newcommand{\BRAMI}{Rice Advanced Material Institute, Rice University, Houston, TX 77005, USA}

\newcommand{\BEQMA}{Extreme Quantum Matter Alliance, Rice University, Houston, TX 77005, USA}

\author{Wentian Zheng}
\thanks{These authors contributed equally to this work.}
\affiliation{\BPhysics}
\affiliation{\BSCI}

\author{Jiawei Hu}
\thanks{These authors contributed equally to this work.}
\affiliation{\BPhysics}

\author{Yuxuan Yao}
\thanks{These authors contributed equally to this work.}
\affiliation{\BPhysics}

\author{Fang Xie}
\affiliation{\BPhysics}

\author{Zihan Wang}
\affiliation{\BPhysics}

\author{Kenji Watanabe}
\affiliation{\BNIMSMaterials}

\author{Takashi Taniguchi}
\affiliation{\BNIMSNano}

\author{Qimiao Si}
\affiliation{\BPhysics}
\affiliation{\BSCI}
\affiliation{\BRAMI}
\affiliation{\BEQMA}

\author{Yonglong Xie}
\email{yx71@rice.edu}
\affiliation{\BPhysics}
\affiliation{\BSCI}
\affiliation{\BRAMI}
\affiliation{\BEQMA}

\setlength{\skip\footins}{12pt plus 2pt minus 3pt}
\let\savedfootnotesize\footnotesize
\renewcommand{\footnotesize}{\fontsize{9}{10.5}\selectfont}
\maketitle
\let\footnotesize\savedfootnotesize

\begin{figure*}[!t]
  \centering
  \figureasset{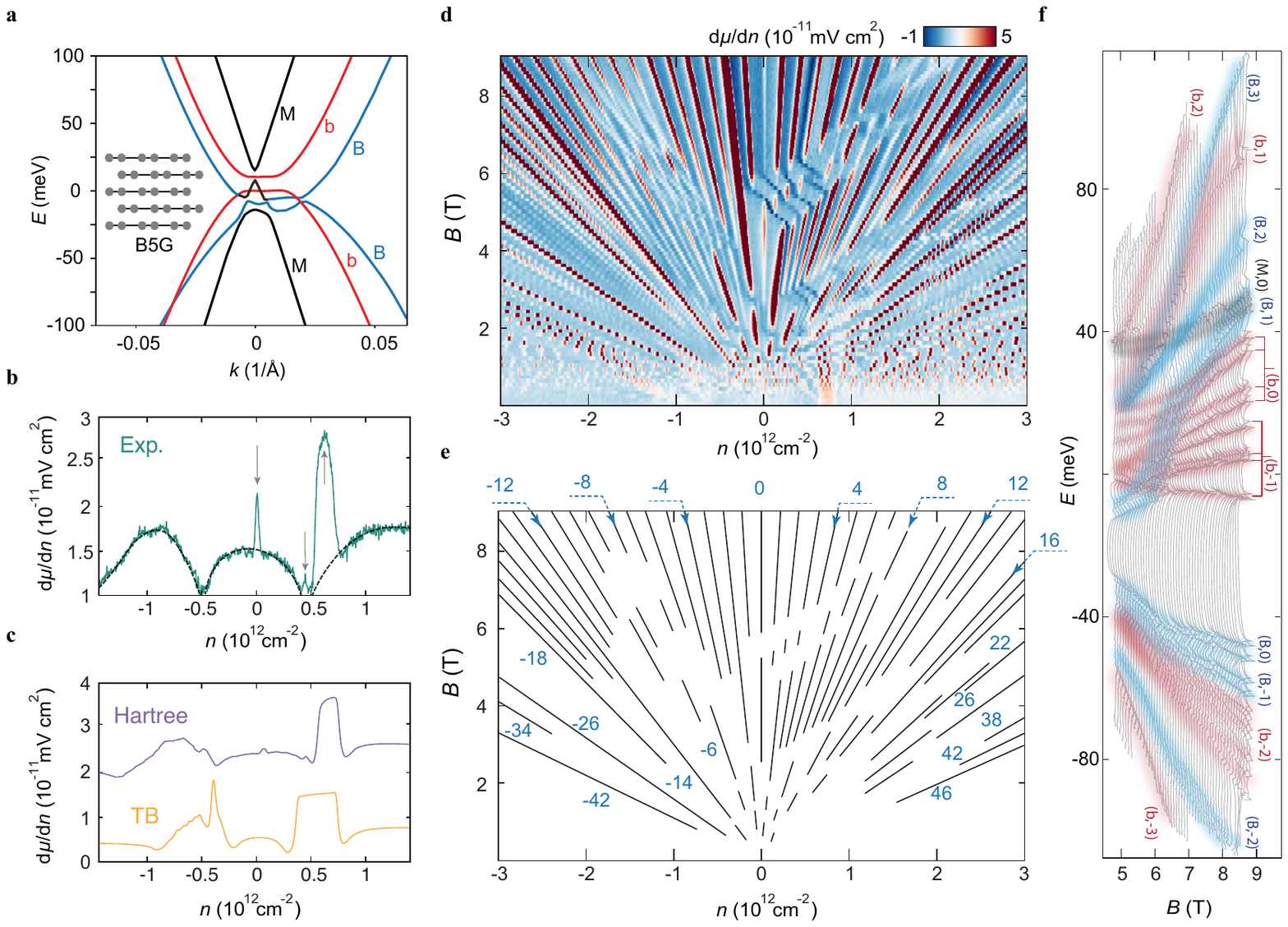}{Figure 1}
  \setlength{\abovecaptionskip}{-8pt}
  \vspace{-6pt}
  \caption{
    \textbf{Sub-band structures and Landau level spectroscopy of B5G.}
    \textbf{a}, Band structure of B5G at zero displacement field and zero doping calculated using a tight-binding model.
    It shows the monolayer-like ($M$, black), light bilayer-like ($b$, red), and heavy bilayer-like ($B$, blue) sub-bands.
    \textbf{b}, Local inverse compressibility as a function of carrier density measured at zero magnetic field.
    Grey arrows mark the three prominent peaks, and dashed curves highlight the three bump-like features.
    \textbf{c}, Calculated inverse compressibility using a tight-binding model (orange) and self-consistent Hartree simulation (purple), respectively.
    The two traces are vertically offset for clarity.
    \textbf{d}, Local inverse compressibility $\dif\mu/\dif n$ map as a function of carrier density and magnetic field.
    \textbf{e}, Wannier diagram highlighting the incompressible states resolved in experimental data.
    The solid lines denote the gapped quantum Hall states and are labeled by their corresponding filling factors.
    \textbf{f}, An experimentally reconstructed many-body Landau level (LL) spectrum. LLs originating from each sub-band are highlighted by different colors, labeled with the subband and orbital LL indices.
  }
  \label{fig:1}
\end{figure*}

\begin{figure*}[t]
  \centering
  \figureasset{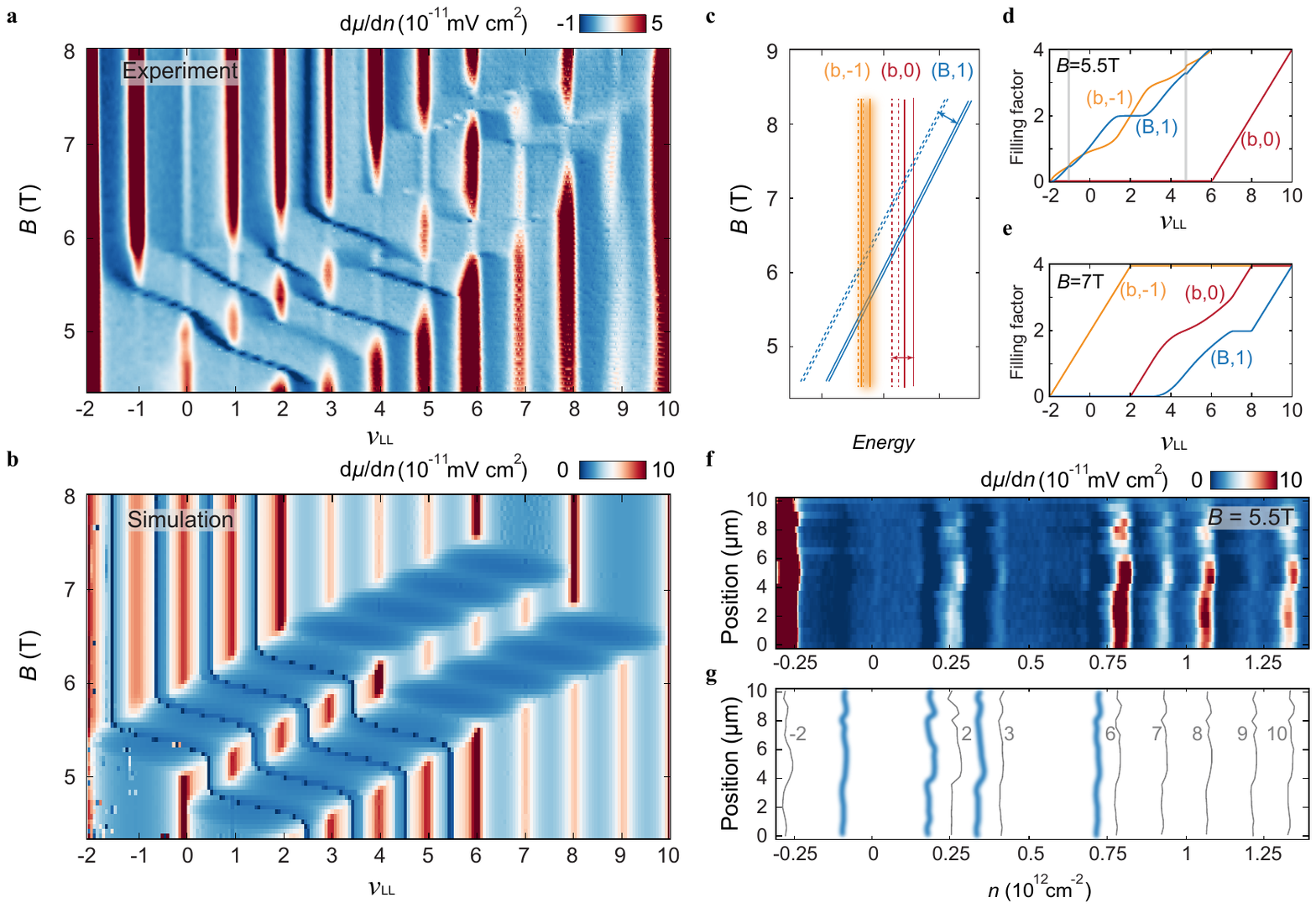}{Figure 2}
  \setlength{\abovecaptionskip}{-8pt}
  \vspace{-10pt}
  \caption{
    \textbf{Negative compressibility staircase as evidence for sub-band phase transitions.}
    \textbf{a}, High-resolution inverse compressibility $\dif\mu/\dif n$ map as a function of Landau filling factor $\nu_{\mathrm{LL}}$ and magnetic field.
    \textbf{b}, Simulated inverse compressibility using the mean-field model described in the main text (see also the Methods).
    \textbf{c}, Single-particle Landau level diagram used in the simulation.
    Solid and dashed lines denote $K$ and $K'$ valley degeneracy, respectively.
    \textbf{d}-\textbf{e}, Population in each sub-band as a function of the total filling factor $\nu_{\mathrm{LL}}$ at $B=5.5\,\mathrm{T}$ (\textbf{d}) and $B=7\,\mathrm{T}$ (\textbf{e}), respectively.
    \textbf{f}, Spatial dependence of local inverse compressibility as a function of carrier density measured at $B=5.5\,\mathrm{T}$ over a $10\,\mu\mathrm{m}$ range.
    \textbf{g}, Trajectories of the carrier density at which quantum Hall states (grey) and phase transitions (blue) appear in (\textbf{f}).
    The number denotes the filling factor of each quantum Hall state.
  }
  \label{fig:2}
\end{figure*}

\textbf{In two-dimensional electron systems, strong Coulomb interactions can spontaneously lift flavor degeneracies such as spin and valley, producing first-order transitions with hysteretic switching of interest for memory devices. The subband degree of freedom in multiband systems may be viewed as a flavor that not only participates in such transitions but also enables control of the hysteretic behavior, as a displacement field can shift the energy difference between subbands. Here, we demonstrate this principle in Bernal-stacked pentalayer graphene, whose low-energy spectrum comprises a monolayer-like Dirac sub-band together with light and heavy bilayer-like sub-bands. Local compressibility measurements resolve a fully symmetry-broken Landau-level spectrum and a sequence of phase transitions that occur when Landau levels of different subbands cross, at which strong interactions redistribute the occupation among flavors. We show that these transitions exhibit hysteretic Hall responses in magnetic field, confirming their first-order character, and that the magnitude and sign of the hysteresis are tuned by the displacement field, as expected from the subband Landau-level ordering determined in compressibility measurements. Our work establishes the subband degree of freedom in multiband systems as a resource for engineering hysteretic switching, and highlights how local thermodynamic measurements can link microscopic insights to macroscopic responses in quantum materials.}

\vspace{3pt}
Strong interactions in systems with a large density of states --- whether from an intrinsically flat band or from Landau levels in a magnetic field --- can spontaneously redistribute the electrons into a subset of their internal degrees of freedom, or flavors, forming flavor-polarized states.
When the manifold of polarized configurations is separated by sufficiently small energy differences, such redistribution can be first-order and exhibit hysteretic switching that can be harnessed for memory applications. This principle underlies the magnetic hysteresis observed in a variety of two-dimensional systems, including GaAs quantum wells~\cite{Cho1998SpinTransitions,Piazza1999FirstOrder,Smet2001Ising,Chen2003Metamorphosis}, moir\'e superlattices~\cite{Sharpe2019EmergentFerromagnetism,Anderson2023MoireGeometry,Liu2023FirstOrder,Waters2024TopologicalFlatBands,Polshyn2020ElectricalSwitching}, and, more recently, graphene multilayers~\cite{Zhou2021HalfQuarterMetals,Han2023OrbitalMultiferroicity,Han2024LargeQAHE,Shi2020ElectronicPhaseSeparation,Chen2023BernalTetralayerAHE,Xie2025TunableFCI,Li2025StackingOrientation,Li2026AnomalousHallCrystals}, where the competing configurations descend from a \emph{single} band with spin and/or valley degeneracy. In the latter two, a displacement field tunes the hysteresis by modifying the width or topology of that single band.
In multiband systems, by contrast, the subband degree of freedom enters as an additional flavor, and the competing electronic configurations may descend from different subbands.
Because the subbands can differ in their layer composition, a displacement field shifts their relative energies, thereby selecting which configurations compete, offering a new strategy for manipulating hysteretic behavior.

Bernal-stacked multilayer graphene naturally realizes this situation. Their low-energy spectrum decomposes into monolayer-like and bilayer-like subbands, whose numbers and effective masses are determined by the thickness~\cite{Koshino2007OrbitalDiamagnetism}.
These subbands are supported on different subsets of layers, so that their relative energies and hybridization are tunable by a displacement field.
In a perpendicular magnetic field, the Landau levels of different subbands disperse at different rates and cross one another within an experimentally accessible range of field~\cite{Koshino2011LandauLevels}.
Transport experiments on Bernal trilayer and tetralayer graphene have indeed resolved resistive features at such crossings that can be tuned by a displacement field~\cite{Stepanov2016TunableSymmetries,Datta2017QuantumHallFerromagnetism,Shi2018TunableLifshitz}.
These features have been attributed to transitions between distinct quantum Hall ferromagnetic states, making Bernal multilayer graphene a promising multiband platform for exploiting the subband degeneracy to control magnetic hysteresis -- though no such hysteresis has yet been reported.

In this work, we report tunable magnetic hysteresis in Bernal-stacked pentalayer graphene (B5G), enabled by interaction-driven redistribution of electrons among isospin-polarized Landau levels of different subbands. The low-energy spectrum of B5G comprises a monolayer-like Dirac sub-band ($M$ band) together with light and heavy bilayer-like sub-bands ($b$ and \textit{B} bands) (Fig.~\ref{fig:1}a), providing a spectrum rich enough that Landau levels of different subbands cross repeatedly in an accessible range of magnetic field, yet simple enough that the competing states at each crossing can be experimentally identified.
Using high-resolution local compressibility measurements with a home-built scanning single-electron transistor (SET) microscope, we resolve fully symmetry-broken quantum Hall states, separated in places by extended regions of negative compressibility that mean-field calculations attribute to a redistribution of electrons between Landau levels of the two bilayer-like subbands.
Transport measurements show that these transitions exhibit magnetic hysteresis in Hall resistance, establishing their first-order character, and that the hysteresis can be tuned by the carrier density and displacement field. Our results establish multiband systems as a versatile setting for engineering tunable phase transitions and magnetic hysteresis.

All devices studied in this work are double-encapsulated B5G with graphite gates.
Local compressibility measurements were performed on two single-gated devices, in which the back-gate voltage tunes both the carrier density and the displacement fields.
Transport measurements were performed on one dual-gated device, in which carrier density and displacement fields can be independently tuned (see Methods).

\vspace{-10pt}
\section*{Many-body Landau level spectroscopy}
\vspace{-10pt}
We begin by examining the local inverse compressibility $\dif\mu/\dif n$ at zero magnetic field, which imprints the subband structure of B5G.
Instead of a single incompressible peak at the charge neutrality point (CNP), we observe three peaks (grey arrows) superimposed on three broad, bump-like features (dashed curves).
To understand these features, we compute the inverse compressibility using a ten-band tight-binding model that assumes a linear potential across the five graphene layers (orange curve in Fig.~\ref{fig:1}c) or layer potentials determined self-consistently within the Hartree approximation (purple curve in Fig.~\ref{fig:1}c) ~\cite{Avetisyan2009ElectricField,Koshino2009}.
We find that only the latter reproduces the salient features of the measured $\dif\mu/\dif n$. Analysis of the self-consistent band structure associates the rightmost peak with the reduced density of states as the chemical potential passes through the gap in the light Bernal-like subband, and the other peaks and broad features with Lifshitz transitions (see Methods and Fig.~\ref{fig:s1} for detailed discussions).
Overall, the excellent agreement between the zero-field data and the self-consistent calculations allows us to constrain the hopping parameters and determine the relative Dirac-point energies of the three subbands, providing a basis for interpreting the Landau-level sequence.

Applying a perpendicular magnetic field quantizes each of these sub-bands into a sequence of Landau levels (LLs), which appear in $\dif\mu/\dif n$ as numerous incompressible peaks dispersing linearly with field (Fig.~\ref{fig:1}d).
Each peak corresponds to a fully gapped quantum Hall state and its slope in the $n$--$B$ plane gives the LL filling factor.
We resolve symmetry-broken states up to $|\nu_{\mathrm{LL}}|=14$ and cyclotron gap up to at least $|\nu_{\mathrm{LL}}|=42$, attesting to the exceptional quality of the sample and the presence of strong electron-electron interaction.
Several of these peaks are interrupted over a finite range of magnetic field and carrier density, as LLs from sub-bands with different Dirac point energies disperse at different rates and cross.
Such crossings have previously been detected as resistive features in longitudinal resistance measurements in Bernal trilayer and tetralayer graphene~\cite{Stepanov2016TunableSymmetries,Datta2017QuantumHallFerromagnetism,Shi2018TunableLifshitz}.

In addition to identifying quantum Hall states, a unique strength of our technique is its ability to reconstruct the many-body LL spectrum shown in Fig.~\ref{fig:1}f (see Methods).
At each magnetic field, the many-body density of states is plotted as a function of energy, allowing a direct visualization of LL dispersion and crossings.
Based on the band structure information inferred from the zero-field analysis, we label the levels by their subbands and LL orbital index ($M$, $b$, or $B$; $N$), with different colors distinguishing the subbands.
This assignment enables us to determine, for each crossing in the measured spectrum, which two levels are brought into coincidence, and hence which states compete at the transitions.

\vspace{-10pt}
\section*{Evidence of first-order phase transitions}
\vspace{-10pt}
Among the interruptions of incompressible peaks in Fig.~\ref{fig:1}d, those near $B=5.5\,\mathrm{T}$ and $n=0.3\times10^{12}\,\mathrm{cm}^{-2}$ stand out because $\dif\mu/\dif n$ is negative over extended regions, while it remains small but positive at other crossings.
Fig.~\ref{fig:2}a shows a high-resolution map of the local inverse compressibility in this region as a function of magnetic field and the LL filling factor $\nu_{\mathrm{LL}}=nh/eB$, where $h$ is Planck's constant and $e$ is the electron charge.
Four pronounced features with $\dif\mu/\dif n<0$ extend through the intervening compressible regions and form a staircase pattern that connects segments of quantum Hall states at successive fillings between $\nu_{\mathrm{LL}}=-2$ and +6.
Similar features were recently observed in magic-angle twisted trilayer graphene and were attributed to interaction-driven phase transitions at which electrons are transferred between nearly degenerate zeroth LLs of the light Dirac and heavy flat sub-bands~\cite{Pierce2025LightHeavy}.
The band structure of B5G also hosts multiple sub-bands.
The experimentally reconstructed LL spectrum in Fig.~\ref{fig:1}f suggests that the observed negative compressibility arises from interaction-induced electron redistribution between the zeroth LL manifold of the light bilayer-like sub-band, comprising $(b,-1)$ and $(b,0)$, and the first dispersive LL $(B,1)$ of the heavy bilayer-like sub-band.

A phenomenological mean-field model that only retains the three LLs involved in the crossings reproduces the salient features of the staircase-like negative compressibility (Fig.~\ref{fig:2}b).
Following previous work~\cite{Zondiner2020Cascade,Pierce2021ChernInsulators,Park2021}, we couple the levels through an on-site Coulomb interaction $U$, which is independent of spin, valley, and subband, and compute the inverse compressibility by minimizing the free energy over the LL occupations at each magnetic field and filling factor (see Methods).
We obtain the best agreement with the single-particle level diagram shown in Fig.~\ref{fig:2}c, which treats $(b,-1)$ as valley-degenerate and includes valley splitting for $(b,0)$ and $(B,1)$.
The calculated occupations show that the intervals of negative $\dif\mu/\dif n$ at $B=5.5\,\mathrm{T}$ coincide with the transfer of electron population  between $(b,-1)$ and $(B,1)$ (Fig.~\ref{fig:2}d).
By contrast, at $B=7\,\mathrm{T}$, the three sets of LLs, $(b,-1)$, $(b,0)$, and $(B,1)$, are filled sequentially, and the negative $\dif\mu/\dif n$ disappear (Fig.~\ref{fig:2}e).
Although the microscopic origin of the differences in valley splitting within each subband remains unknown, the remarkable correspondence establishes that the subband degree of freedom, like spin and valley, plays a key role in the electronic phase transitions.

Spatial-dependent compressibility measurements over an extended range of $10\,\mu\mathrm{m}$ demonstrate the remarkable reproducibility of these subband phase transitions (Fig.~\ref{fig:2}f).
The negative-compressibility features shift only slightly and generally follow the spatial variation of the nearby quantum Hall states (Fig.~\ref{fig:2}g). These correlated shifts are consistent with local charge disorder producing a common density offset. The nearly fixed positions of the transitions relative to the neighboring quantum Hall states further support their assignment to charge redistribution among the crossing LLs. In magic-angle twisted trilayer graphene, by contrast, the negative-compressibility features shift drastically relative to neighboring quantum Hall states, due to additional spatial variations in twist angle, strain, or other microscopic parameters~\cite{Pierce2025LightHeavy}.
We also observe similar negative compressibility features in a second B5G device aligned with hBN, establishing that these phase transitions are reproducible salient features and remain intact in the presence of a moir\'e potential (Fig.~\ref{fig:s2}).
Together, these observations suggest that the robustness of these transitions revealed by local compressibility measurements may leave a detectable signature in macroscopic transport.

\begin{figure*}[t]
  \centering
  \figureasset{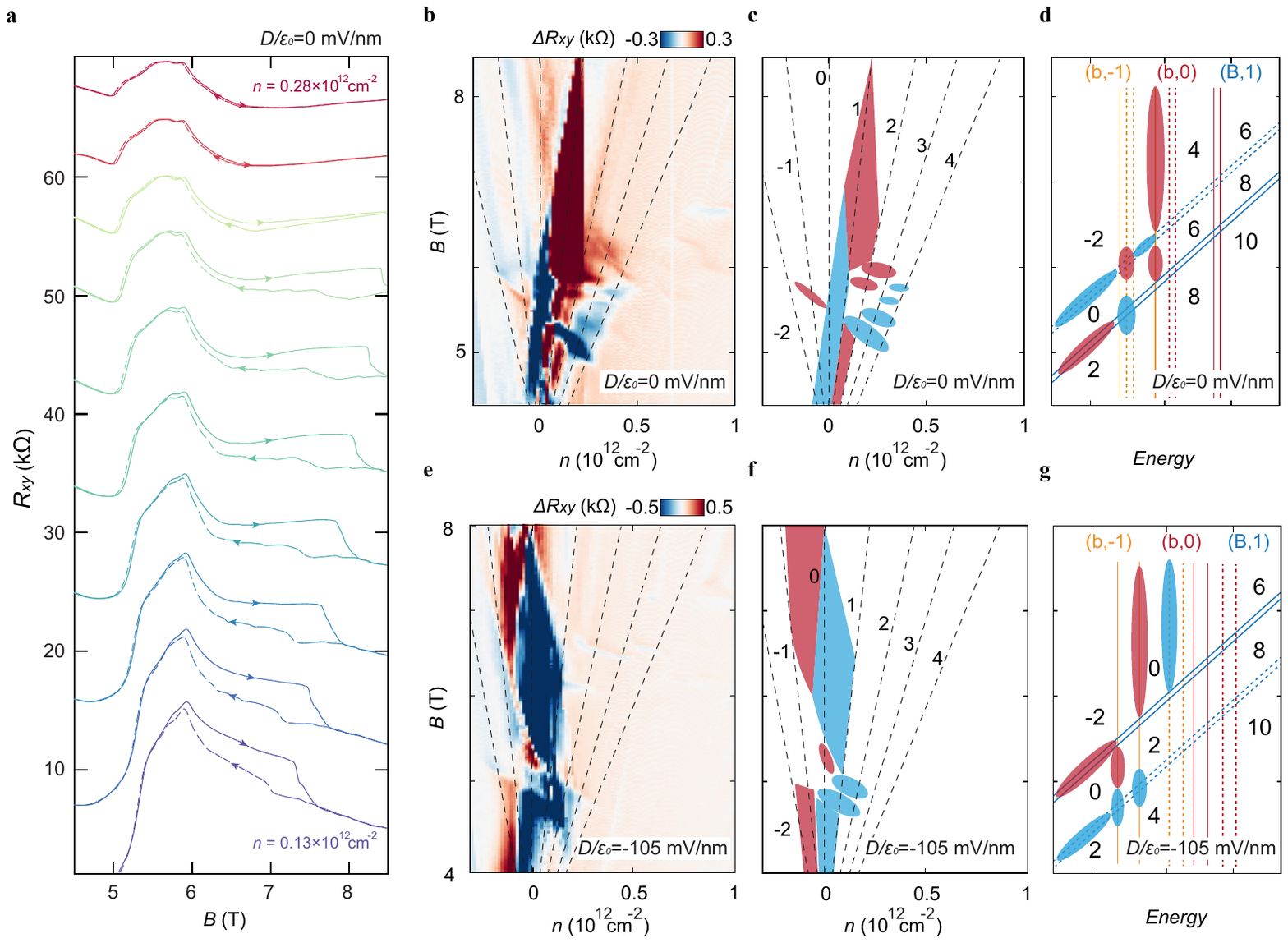}{Figure 3}
  \setlength{\abovecaptionskip}{-8pt}
  \caption{
    \textbf{Displacement-field-tuned magnetic hysteresis.}
    \textbf{a}, Hall resistance $R_{xy}$ as a function of magnetic field measured at zero displacement field at selected carrier densities from $0.13\times10^{12}\,\mathrm{cm}^{-2}$ to $0.28\times10^{12}\,\mathrm{cm}^{-2}$.
    Solid and dashed lines denote increasing- and decreasing-field sweeps, respectively.
    Traces at different carrier densities are vertically offset for clarity.
    \textbf{b}, Map of hysteresis $\Delta R_{xy}$ (defined as the up-sweep minus the down-sweep) at $D/\varepsilon_0=0\,\mathrm{mV}\,\mathrm{nm}^{-1}$ as a function of magnetic fields and carrier density.
    Dashed lines mark the dispersion of the quantum Hall states.
    \textbf{c}, Schematic of the positive (red) and negative (blue) hysteresis regions observed in (\textbf{b}).
    The same quantum Hall states are marked as dashed lines and labeled by their filling factor.
    \textbf{d}, Schematic single-particle Landau level diagrams at $D/\varepsilon_0=0\,\mathrm{mV}\,\mathrm{nm}^{-1}$, showing the relative positions and internal splitting of the $(b,-1)$, $(b,0)$, and $(B,1)$ manifolds.
    Numbers denote the filling factors of each gapped state.
    Solid and dashed lines denote $K$ and $K'$ valley degeneracy, respectively.
    \textbf{e}, Map of hysteresis $\Delta R_{xy}$ at $D/\varepsilon_0=-105\,\mathrm{mV}\,\mathrm{nm}^{-1}$.
    \textbf{f}, Schematic of the positive (red) and negative (blue) hysteresis regions observed in (\textbf{e}).
    \textbf{g}, Schematic single-particle Landau level diagrams at $D/\varepsilon_0=-105\,\mathrm{mV}\,\mathrm{nm}^{-1}$.
  }
  \label{fig:3}
\end{figure*}

\vspace{-10pt}
\section*{Tunable magnetic hysteresis}
\vspace{-10pt}
To probe the transport signature of the sub-band phase transitions and test whether they are first-order, we measure the longitudinal and Hall resistances of a dual-gated B5G device.
Consistent with previous studies~\cite{Stepanov2016TunableSymmetries,Datta2017QuantumHallFerromagnetism,Shi2018TunableLifshitz}, maps of longitudinal conductance versus carrier density $n$ and magnetic field $B$ display features associated with LL crossings (Fig.~\ref{fig:s3}).
To look for hysteresis in this region, we first compare $R_{xy}$ measured on increasing and decreasing magnetic-field sweeps at zero displacement field.
Figure.~\ref{fig:3}a shows the up-sweep (solid) and down-sweep (dashed) traces at representative carrier densities from $n=0.13\times10^{12}\,\mathrm{cm}^{-2}$ to $n=0.28\times10^{12}\,\mathrm{cm}^{-2}$, with differences of several hundred ohms to a few kilo-ohms where the traces separate.
Mapping this difference, $\Delta R_{xy}$, across the $n$--$B$ plane reveals an intricate pattern of positive and negative regions (Fig.~\ref{fig:3}b), illustrating how both the carrier density and magnetic field can be used to tune the hysteretic Hall response. Fig.~\ref{fig:3}c summarizes this pattern schematically, with the LL trajectories overlaid as dashed lines.

Using the LL diagram inferred from the compressibility measurements (Fig.~\ref{fig:3}d), we interpret the $\Delta R_{xy}$ pattern by considering how the orbital magnetization is expected to evolve as different levels are filled.
The strongest feature occurs at $1<\nu_{\mathrm{LL}}<2$ and $6\,\mathrm{T}<B<8.5\,\mathrm{T}$ and likely corresponds to filling a spin- and valley-polarized level in the $(b,-1)$ manifold, echoing the hysteresis observed at transitions between symmetry-broken quantum Hall states in GaAs~\cite{Piazza1999FirstOrder}. A similar link between hysteresis and polarization appears on the low-field side of the subband phase transitions $B<5.5\,\mathrm{T}$.
The twofold spin-degenerate $K'$ and $K$ levels in the $(B,1)$ manifold carry opposite orbital magnetization, so filling them is expected to change the total orbital magnetization in opposite directions.
This is consistent with the opposite-sign hysteretic features observed at $-2<\nu_{\mathrm{LL}}<0$ and $0<\nu_{\mathrm{LL}}<2$, where the $(B,1,K')$ and $(B,1,K)$ levels are filled, respectively.
Between $B=5.5\,\mathrm{T}$ and $B=6.5\,\mathrm{T}$, the $\Delta R_{xy}$ tracks the regions of the observed negative compressibility, supporting the first-order character of the subband phase transitions.
Transitions from $(B,1,K')$ to different flavors of $(b,-1)$ produce multiple positive hysteresis signals, while transitions from $(B,1,K)$ lead to hysteresis of opposite sign.
These findings establish the subband transitions as an additional means of engineering magnetic hysteresis.

A displacement field changes the electrostatic potential across the five graphene layers, and is expected to modify the LL structures and the resulting subband phase transitions and magnetic hysteresis.
Fig.~\ref{fig:3}e shows a representative hysteresis map at $D/\varepsilon_0=-105\,\mathrm{mV}\,\mathrm{nm}^{-1}$.
Compared to $D=0$, we observe several notable differences.
First, for $6\,\mathrm{T}<B<8\,\mathrm{T}$, the feature associated with filling $(b,-1,K')$ becomes negative relative to ($D=0$), and a positive feature appears where $(b,-1,K)$ fills.
The two features are consistent with a change in the valley splitting of $(b,-1)$ and with the opposite orbital magnetic moments of the two valleys.
Second, both low-field features reverse sign, suggesting that the ($K'$) and ($K$) levels in $(B,1)$ exchange their energy order.
Third, near the subband phase transitions, the lower boundary of the hysteretic region shifts to lower magnetic fields, consistent with the change in the $(b,-1)$ valley splitting inferred from the high-field response. The schematic finite-$D$ LL diagram summarizes these assignments (Fig.~\ref{fig:3}g).
Because the dual-gate geometry precludes SET compressibility measurements on this device, we construct the LL diagram by shifting the $D=0$ levels according to the evolution of the longitudinal conductance maps with $D$~\cite{Stepanov2016TunableSymmetries} (Fig.~\ref{fig:s3}). Regardless, these measurements demonstrate that the displacement field provides rich electrostatic control over hysteresis associated with electron repopulation between subbands.

\vspace{-10pt}
\section*{Discussion}
\vspace{-10pt}
The joint thermodynamic and transport experiments presented here demonstrate how the sub-band degree of freedom in a multiband system can give rise to first-order phase transitions and provide a means of controlling the associated magnetic hysteresis.
In B5G, a displacement field tunes the onset, magnitude, and sign of this hysteresis, consistent with a change in the relative energies of the LLs from different sub-bands.
An open question is what favors first-order transitions at particular LL crossings, given that extended negative compressibility is confined to the electron-doped side near charge neutrality.
Understanding this selectivity could guide the search for analogous transitions in other multiband systems, including those with closely spaced flat bands at zero magnetic fields.

\bibliographystyle{apsrev4-1}
\bibliography{references}

\vspace{-10pt}
\section*{Acknowledgments}
\vspace{-10pt}
This work was primarily supported by Robert A. Welch Foundation Grant No. C-2219 (device fabrication) and the NSF CAREER DMR-2339623 (manuscript writing).
Additional support comes from a David and Lucile Packard Fellowship for Science and Engineering and the Scialog grant \#SA-QMI-2025-081a from Research Corporation for Science Advancement and the Brinson Foundation. W.Z. acknowledges the Peter M \& Ruth L Nicholas Postdoctoral Fellowship in Nanotechnology from the Smalley-Curl Institute at Rice University. Y.Y. was partially supported by the Nettie S. Autrey Fellowship. F.X. and Q.S. was supported by the U.S. Department of Energy, Office of Science, Basic Energy Sciences, under Award No. DE-SC0018197. K.W. and T.T. acknowledge support from the CREST (JPMJCR24A5), JST and World Premier International Research Center Initiative (WPI), MEXT, Japan.
We disclose the use of GPT-5.6 Sol to generate the code for the self-consistent Hartree simulation and to polish the manuscript.

\vspace{-10pt}
\section*{Author contributions}
\vspace{-10pt}
W.Z., J.H., Y.Y., and Y.X. conceived and designed the experiment.
W.Z., J.H., and Y.Y. built the scanning SET microscope and performed the local compressibility measurements.
Y.Y. fabricated the samples and performed the transport measurements with assistance from J.H.,W.Z. and Z.W.
F.X. constructed the tight-binding model under the supervision of Q.S.
W.Z. and J.H. performed the numerical simulation with input from F.X.
K.W. and T.T. provided the hBN crystals.
Y.X. and W.Z. wrote the manuscript with input from all authors.

\vspace{-10pt}
\section*{Data availability}
\vspace{-10pt}
The data that supports the findings of this study are available from the corresponding authors upon reasonable request.

\vspace{-10pt}
\section*{Code availability}
\vspace{-10pt}
The codes that support the findings of this study are available from the corresponding authors upon reasonable request.

\vspace{-10pt}
\section*{Methods}
\begingroup
\setlength{\parskip}{4pt}
\vspace{-10pt}
\textbf{Sample Preparation.} Pentalayer graphene, graphite, and hBN flakes were mechanically exfoliated onto SiO$_2$/Si substrates and identified under an optical microscope. The stacking order of the pentalayer graphene was determined using wide-field infrared imaging and further confirmed by Raman spectroscopy.
The van der Waals heterostructures were assembled using a standard dry transfer technique: top hBN and pentalayer graphene were picked up with a polypropylene carbonate film and transferred onto a prepared bottom hBN with a graphite back gate to form the single-gated devices D1 and D2.
In D2, we carefully aligned the top hBN crystal with the pentalayer graphene to form a moir\'e superlattice.
For the dual-gated device D3, an additional hBN layer and a graphite top gate were subsequently transferred onto it.
The stack was then patterned into a Hall-bar geometry using e-beam lithography and reactive-ion etching, followed by Cr/Au deposition to form electrical contacts to the sample and gate electrodes.

\textbf{Local compressibility measurements.} Compressibility measurements were carried out in a home-made \({}^{3}\mathrm{He}\) SET microscope system equipped with a vector magnet, operating at a base temperature of about $300\,\mathrm{mK}$.
During measurements, the SET tip was brought to a height of about $100\,\mathrm{nm}$ above the sample surface. We modulate the sample and back-gate voltages, $V_{\mathrm{2D,ac}}$ and $V_{\mathrm{bg,ac}}$, at $107.177\,\mathrm{Hz}$ and $97.177\,\mathrm{Hz}$, respectively, and measure the a.c. response $I_{\mathrm{2D,ac}}$, $I_{\mathrm{bg,ac}}$ through the SET using standard lock-in techniques.
The inverse compressibility was calculated by \[\frac{\dif\mu}{\dif n}=\frac{e^2(I_{\mathrm{bg,ac}}/V_{\mathrm{bg,ac}})}{C_b(I_{\mathrm{2D,ac}}/V_{\mathrm{2D,ac}})}\]
where $C_b$ is the backgate capacitance per area, determined from the Landau fan diagram.
To hold the SET response fixed, an analog proportional-integral-derivative controller is used to maintain the d.c. SET current at a fixed value by controlling the d.c. voltage of the sample and back gate relative to the tip.
The corresponding change in d.c. voltage provides a direct measure of $\mu(n)$.

\textbf{Transport measurements.} Transport measurements were performed in an Oxford Proteox dilution refrigerator at a base temperature of around $10\,\mathrm{mK}$.
Longitudinal and transverse Hall resistance $R_{xx}$ and $R_{xy}$ were measured using standard lock-in techniques with an a.c. current bias of $100\,\mathrm{nA}$ at a frequency of $7.777\,\mathrm{Hz}$.
The carrier density $n=(C_tV_t+C_bV_b)/e$ and the displacement field $D/\varepsilon_0=(C_tV_t-C_bV_b)/2$ were tuned by sweeping $V_t$ and $V_b$ using voltage sources; here, $C_t$ and $C_b$ are the top-gate and back-gate capacitances per unit area, respectively, calibrated from the Landau fan diagram.

\textbf{Tight-binding and self-consistent Hartree calculations.} The zero-field inverse compressibility was first calculated using a ten-band continuum tight-binding Hamiltonian $H_0(\mathbf{k})$, adopting the hopping parameters $\gamma_0$--$\gamma_5$ and on-site energy shifts $\delta_1$ at the dimer sites and $\delta_2$ in the outermost layers used in previous work\cite{Koshino2011LandauLevels}.
To account for mirror-symmetry breaking and electrostatic screening between layers in our single-gate device geometry, we employ a self-consistent Hartree approximation in which the electrostatic potential in each layer is determined iteratively.
We begin an iteration process as follows: at a given carrier density $n$, considering electrostatic potential of each layer $\{U_1,U_2,\ldots,U_5\}$, the Hamiltonian is modified to $H(\mathbf{k})=H_0(\mathbf{k})+\operatorname{diag}(U_1,U_2,\ldots,U_5)$.
The electron density of each layer $\{n_1,n_2,\ldots,n_5\}$ can be calculated from each eigenstate $\lvert E_{\mathbf{k}}\rangle$ weight $w(\mathbf{k})$ on each layer after diagonalization of $H(\mathbf{k})$.
The chemical potential $\mu$ is then constrained by $4\sum\int \sigma(E(\mathbf{k})-\mu)-\sum_i n_i=n$, where $\sigma$ is the Gaussian-broadened occupation function at finite temperature. A new set of $\{n_1^*,n_2^*,\ldots,n_5^*\}$ can be derived out as $n_i^*=4\sum\int w(\mathbf{k})\sigma(E(\mathbf{k})-\mu)-n_i$.
Then each electrostatic potential is consequently updated to a new value as: $U_{i+1}^*-U_i^*=edE_{i,i+1}$ where $E_{i,i+1}$ denotes the interlayer electrostatic field determined from the arrangement of $n_i^*$ through Gauss's law, and $d$ is the interlayer spacing.
This procedure is iterated until the electrostatic potentials converge, yielding the self-consistent band structure and chemical potential as functions of carrier density.
We find good agreement with experimental data in Fig.~\ref{fig:1}c using the following hopping parameters and energy shifts: $\gamma_0=3.00\,\mathrm{eV}$, $\gamma_1=0.37\,\mathrm{eV}$, $\gamma_2=-0.019\,\mathrm{eV}$, $\gamma_3=0.30\,\mathrm{eV}$, $\gamma_4=0.08\,\mathrm{eV}$, $\gamma_5=0.053\,\mathrm{eV}$, $\delta_1=0.044\,\mathrm{eV}$, $\delta_2=6.71\,\mathrm{meV}$.
The Gaussian energy broadening is set to $25\,\mu\mathrm{eV}$, corresponding to $300\,\mathrm{mK}$, the experimental temperature at which the compressibility measurements are performed.
Finally, we choose the interlayer spacing $d=0.335\,\mathrm{nm}$ and the effective relative permittivity $\varepsilon_r=4$.

\textbf{Reconstruction of Landau level spectrum.} The Landau-level spectrum in Fig.~\ref{fig:1}f was reconstructed from the compressibility map in Fig.~\ref{fig:1}d.
At each magnetic field $B$, we numerically integrated $\dif\mu/\dif n$ with respect to carrier density $n$ to obtain the chemical potential $\mu(n)$.
Using $\mu$ as the energy coordinate, we then plotted the density of states $\dif n/\dif\mu$ and arranged the traces at their corresponding magnetic fields to visualize the Landau-level spectrum.
We note here that interaction-induced negative compressibility regions complicate this reconstruction and can cause neighboring Landau-level features to overlap.
To improve their visibility, each affected inverse compressibility trace was shifted upward by a constant before integration, and the displayed $\dif n/\dif\mu$ was set to zero in regions where the original $\dif\mu/\dif n$ was negative.
Although this adjustment increases the magnitude of the reconstructed chemical potential, the resulting values remain comparable to those obtained from d.c. chemical potential measurements.

\textbf{Mean-field simulation of compressibility.} To understand the negative compressibility between Landau levels, we perform mean-field calculations retaining the three LL manifolds involved in the transitions.
To compute the inverse compressibility at a given value of $\nu$ and $B$, we begin with a set of trial fillings for each degeneracy $\nu_i$ constrained by $\nu=\sum\nu_i$ and compute the Gibbs free energy $G=G_k+G_U$, with contributions from the kinetic energy $G_k$ and interactions $G_U$, respectively.
We model each Landau level with a Gaussian-broadened density of states $D(E)$ of bandwidth $W$, and obtain the filling factor using $\nu_i=\int D(E)\,\dif E$.
The kinetic and interaction terms are given by $G_k=\int E D(E)\,\dif E$ and $G_U=U\sum_{i,j}\nu_i\nu_j-J\sum_i\nu_i^2$, where $U$ and $J$ denote the coupling strength of Coulomb repulsion and exchange interaction, respectively, within the mean-field approximation.
For a given filling factor $\nu$, we determine $\nu_i$ iteratively by minimizing $G$ via gradient descent and compute the inverse compressibility using $\frac{\dif\mu}{\dif n}=\frac{h^2}{e^2B^2}\frac{\partial^2 G}{\partial\nu^2}$.
In Fig.~\ref{fig:2}b, we find good agreement with our experimental data when $W=0.3\,\mathrm{meV}$, the coupling strength $U/W=3$, $J=0$, and the valley splitting within $(b,0)$ and $(B,1)$ was set to $1\,\mathrm{meV}$ and $3\,\mathrm{meV}$.

\textbf{Displacement-field-dependent transport.}
To construct the finite-$D$ LL diagram used to interpret the Hall hysteresis, we measure the four-terminal longitudinal conductance $G_{xx}$ near the subband phase transitions at representative displacement fields.
While these measurements cannot distinguish between subband phase transitions with negative compressibility and standard LL crossings with large positive compressibility, the sequence of the neighboring quantum Hall states allows us to constrain the LL diagram as a function of $D$.
At $D=0$, the quantum Hall features are weak and diffuse within the region bounded by the gray dashed lines in Fig.~\ref{fig:s3}a.
In the LL scheme inferred from compressibility (Fig.~\ref{fig:2}c), this region corresponds to crossings among the $(b,-1)$, $(b,0)$, and $(B,1)$ levels.
The evolution of these boundaries with $D$ constrains the relative energies of the crossing levels.
At $D/\varepsilon_0=61\,\mathrm{mV\,nm^{-1}}$, the lower boundary moves to lower magnetic fields and the upper boundary moves to higher magnetic fields, widening the crossing region (Fig.~\ref{fig:s3}b).
The $\nu_{\mathrm{LL}}=4$ state provides a separate constraint on the valley gaps.
In the inferred LL scheme, this state probes the $(b,-1)$ gap below the crossing region and the $(b,0)$ gap above it.
At $D=0$, the state is unresolved on the low-field side and well resolved on the high-field side, consistent with a smaller valley gap in $(b,-1)$ than in $(b,0)$.
The low-field state emerges at $D/\varepsilon_0=61$ and $-50\,\mathrm{mV\,nm^{-1}}$ and becomes more pronounced as $D/\varepsilon_0$ decreases from $-50$ to $-78\,\mathrm{mV\,nm^{-1}}$, consistent with an increased valley gap in $(b,-1)$.
At $D/\varepsilon_0=-78\,\mathrm{mV\,nm^{-1}}$, the high-field state disappears, suggesting that the $(b,0)$ valley gap is reduced (Fig.~\ref{fig:s3}b--d).
The finite-$D$ LL diagrams in Fig.~\ref{fig:s3}e and Fig.~\ref{fig:s3}f summarize the evolution inferred from the crossing boundaries and the $\nu_{\mathrm{LL}}=4$ state.
This evolution is consistent with the displacement-field dependence of the LL spectrum reported in Bernal trilayer graphene~\cite{Stepanov2016TunableSymmetries}. 

\par
\endgroup

\clearpage
\onecolumngrid
\setcounter{figure}{0}
\renewcommand{\thefigure}{S\arabic{figure}}
\renewcommand{\theHfigure}{supp.\arabic{figure}}

\begin{figure}[!htbp]
  \centering
  \figureasset{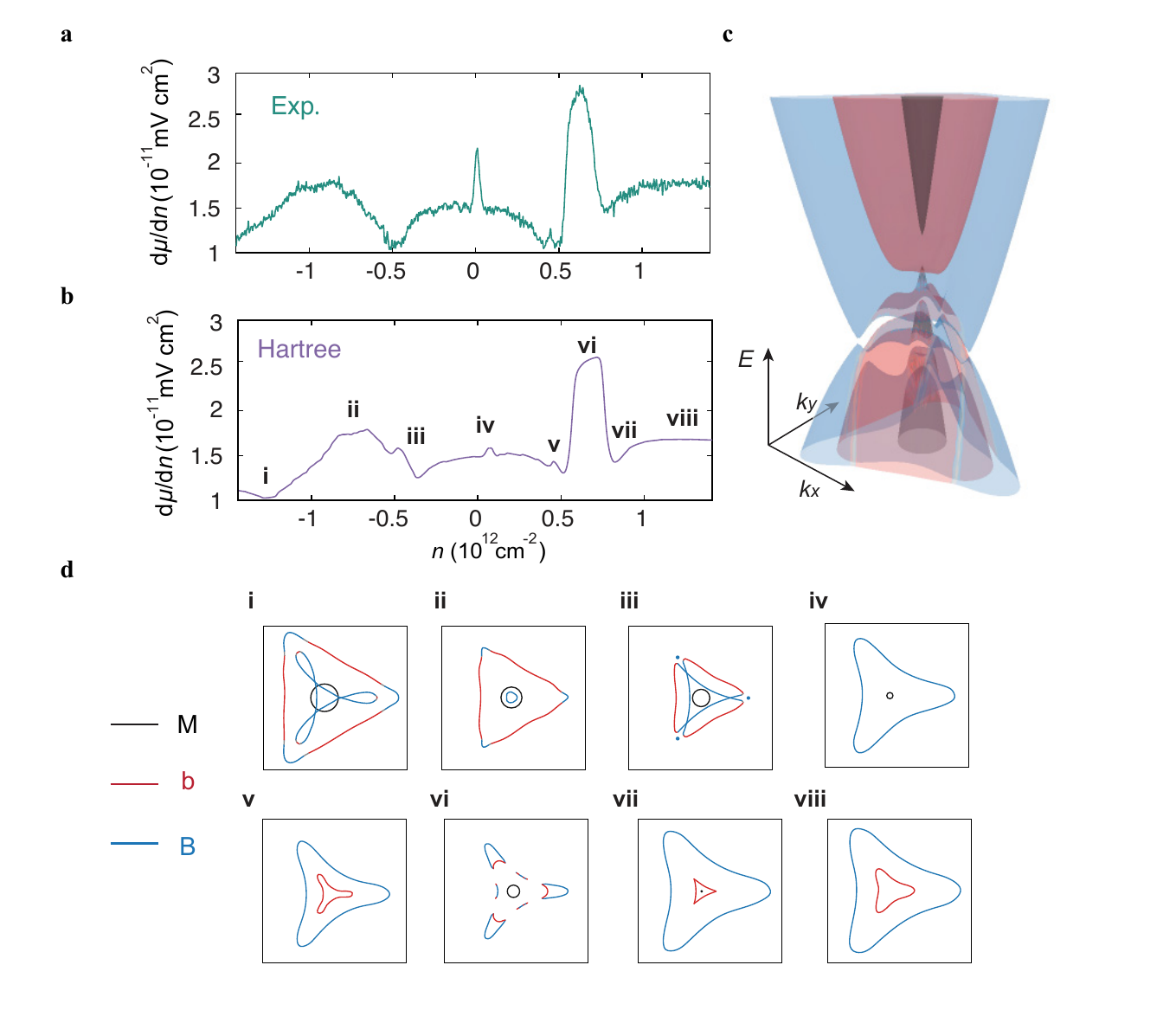}{Figure S1}
  \setlength{\abovecaptionskip}{-8pt}
  \caption{
    \textbf{Fermiology as a function of carrier density.}
    \textbf{a}, Local compressibility measured with the scanning SET, same as Fig.~\ref{fig:1}b.
    \textbf{b}, Simulated compressibility using self-consistent Hartree approximation.
    \textbf{c}, Band structure of B5G at zero displacement field and zero doping, near the $K$ point.
    \textbf{d}, Fermi contours at representative carrier densities marked by Roman numerals in (\textbf{b}).
    Black, red, and blue colors denote the monolayer-like ($M$), light Bernal-like ($b$), and heavy Bernal-like ($B$) subbands, respectively.
  }
  \label{fig:s1}
\end{figure}

\begin{figure}[!htbp]
  \centering
  \figureasset{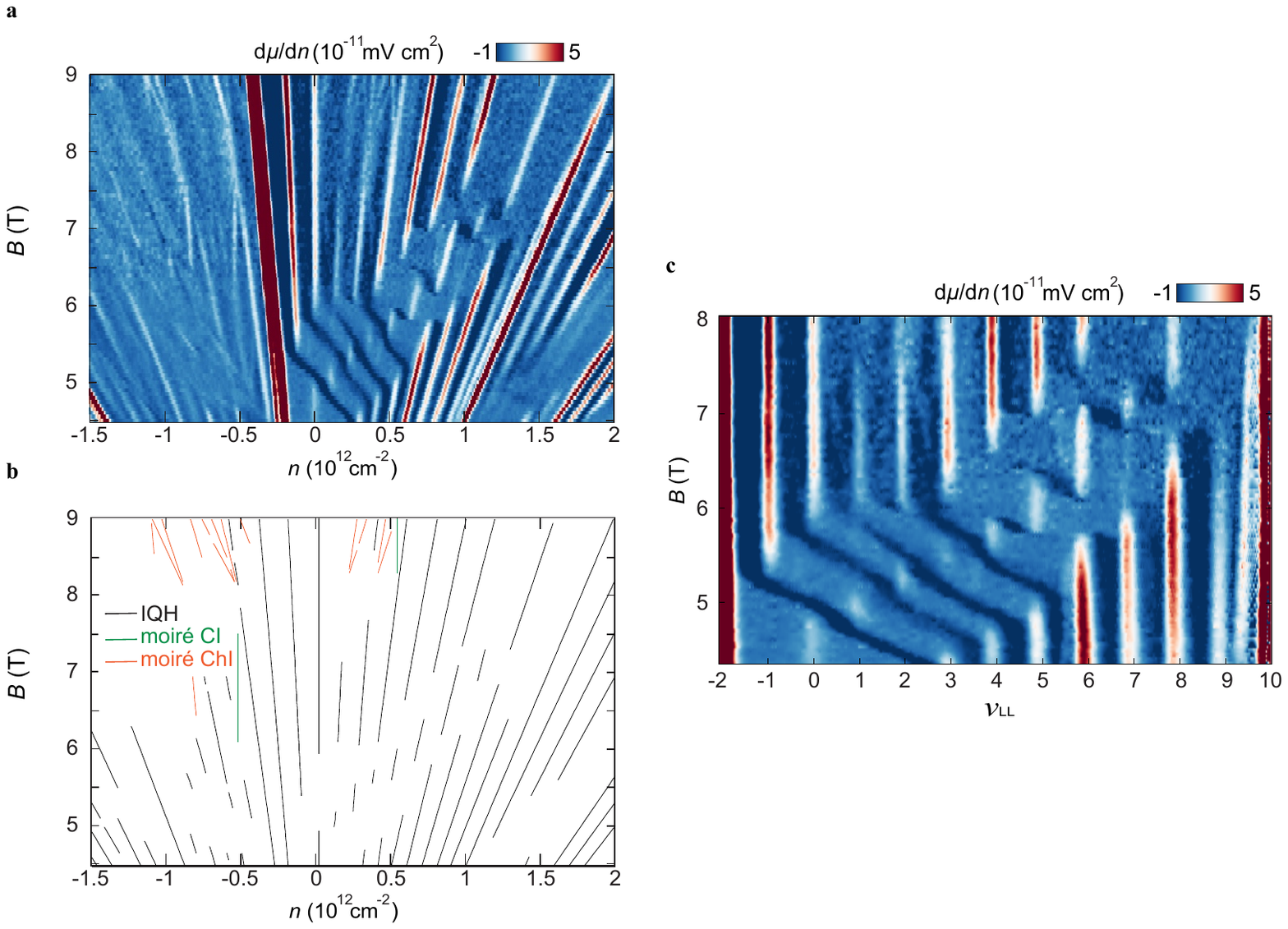}{Figure S2}
  \setlength{\abovecaptionskip}{-8pt}
  \caption{
    \textbf{Negative-compressibility staircase in another B5G device aligned with the hBN (D2).}
    \textbf{a}, Local inverse compressibility $\dif\mu/\dif n$ as a function of carrier density and magnetic field.
    \textbf{b}, Wannier diagram identifying the incompressible states observed in (\textbf{a}).
    Black lines correspond to IQH states.
    Green and orange lines correspond to moir\'e correlated insulators (CIs), and moir\'e Chern insulators (ChIs) with a non-zero filling of the moir\'e unit cell, respectively.
    \textbf{c}, Zoomed-in near the negative-compressibility staircase plotted as a function of Landau filling factor $\nu_{\mathrm{LL}}$ and magnetic field, showing the robustness of the phase transition to the presence of a moir\'e potential.
  }
  \label{fig:s2}
\end{figure}

\begin{figure}[!htbp]
  \centering
  \figureasset{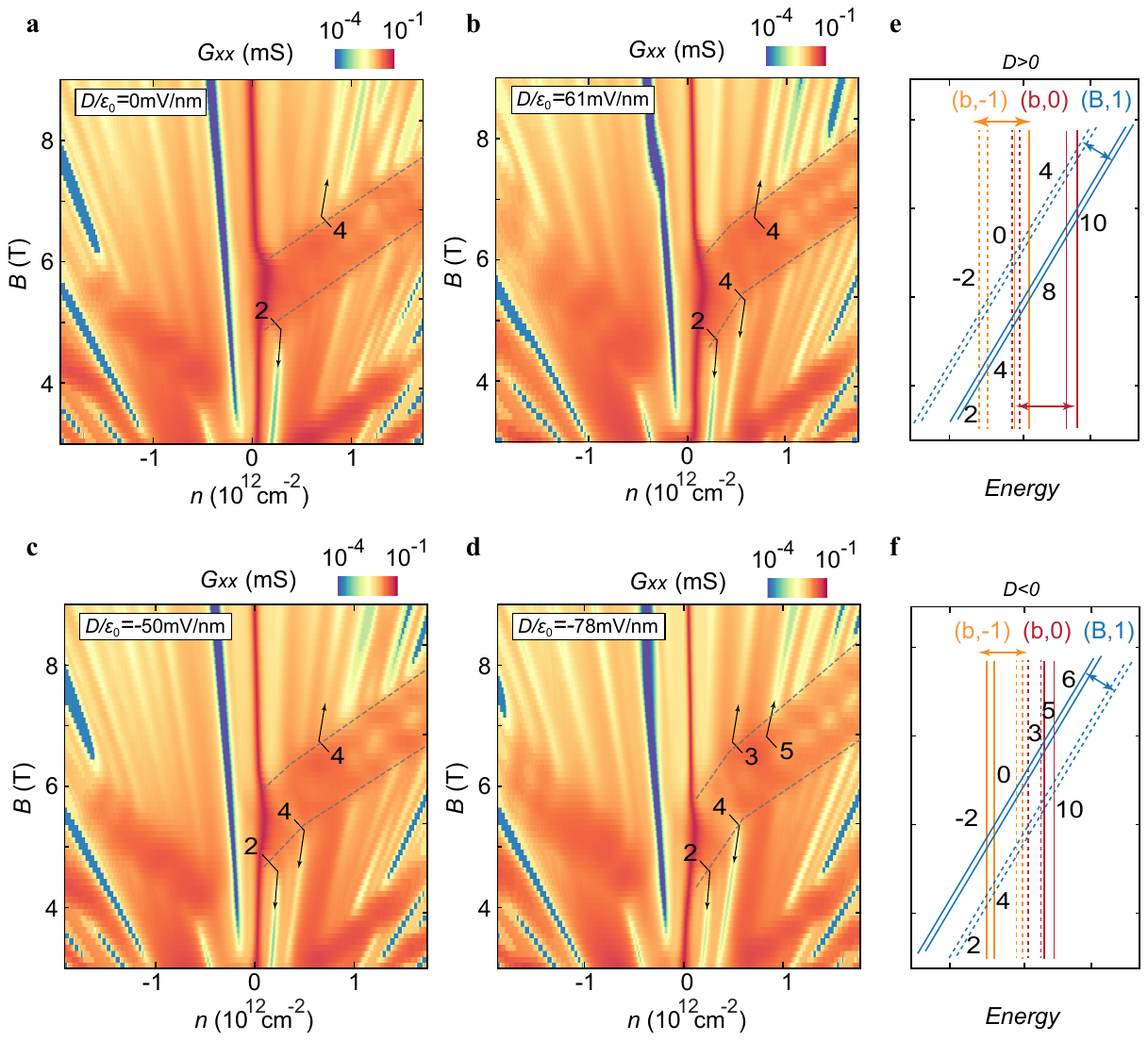}{Figure S3}
  \setlength{\abovecaptionskip}{-8pt}
  \caption{
    \textbf{Displacement-field evolution of the phase transitions.} \textbf{a}-\textbf{d}, Longitudinal conductance $G_{xx}$ as a function of carrier density and magnetic field at displacement field $D=0$ (\textbf{a}), $D/\varepsilon_0=61\,\mathrm{mV}\,\mathrm{nm}^{-1}$ (\textbf{b}), $D/\varepsilon_0=-50\,\mathrm{mV}\,\mathrm{nm}^{-1}$ (\textbf{c}), $D/\varepsilon_0=-78\,\mathrm{mV}\,\mathrm{nm}^{-1}$ (\textbf{d}). $G_{xx}$ is calculated by $G_{xx}=R_{xx}/(R_{xx}^2+R_{xy}^2)$.
    Dashed lines mark the regions of LL crossings of sub-bands; numbers denote the filling factor of each quantum Hall state.
    \textbf{e}-\textbf{f}, Schematic single-particle Landau level diagrams showing the relative positions and internal splitting of the $(b,-1)$, $(b,0)$, and $(B,1)$ manifolds for $D/\varepsilon_0>0$ (\textbf{e}) and $D/\varepsilon_0<0$ (\textbf{f}).
    Numbers denote the filling factors of each gapped state.
    Solid and dashed lines denote $K$ and $K'$ valley degeneracy, respectively.
  }
  \label{fig:s3}
\end{figure}

\end{document}